\documentclass[sigconf,nonacm]{acmart}
\usepackage{graphicx}
\usepackage{listings}
\usepackage{inputenc}
\usepackage{xcolor}
\AtBeginDocument{%
  \providecommand\BibTeX{{%
  \normalfont 
  \kern-0.5em{\scshape i\kern-0.25em b}\kern-0.8em\TeX}
  }}

\usepackage{natbib}

\AtBeginDocument{%
 \providecommand\BibTeX{{%
 \normalfont B\kern-0.5em{\scshape i\kern-0.25em b}\kern-0.8em\TeX}}}

\setcopyright{acmlicensed}
\copyrightyear{2024}
\acmYear{2024}
\acmDOI{XXXXXXX.XXXXXXX}

\acmConference[Conference acronym 'XX]{Make sure to enter the correct
 conference title from your rights confirmation email}{June 03--05,
 2018}{Woodstock, NY}
\acmISBN{978-1-4503-XXXX-X/18/06}

\begin{document}

\title{Climate AI for Corporate Decarbonization Metrics Extraction}



\author{Aditya Dave}
\email{aditya.dave@blackrock.com}
\orcid{0000-0001-6021-0243}
\affiliation{%
  \institution{BlackRock, Inc.}
  \city{New York}
  \state{NY}
  \country{USA}
}

\author{Mengchen Zhu}
\email{mengchen.zhu@blackrock.com}
\affiliation{%
  \institution{BlackRock, Inc.}
  \city{New York}
  \state{NY}
  \country{USA}}

\author{Dapeng Hu}
\email{dapeng.hu@blackrock.com}
\affiliation{%
  \institution{BlackRock, Inc.}
  \city{New York}
  \state{NY}
  \country{USA}}

  \author{Sachin Tiwari}
  \email{sachin.tiwari@blackrock.com}
  \affiliation{%
    \institution{BlackRock, Inc.}
    \city{Gurgaon}
    \state{Haryana}
    \country{IND}}
%

\begin{abstract}

  Corporate Greenhouse Gas (GHG) emission targets are important metrics in sustainable investing \cite{dahlmann2019managing,krabbe2015aligning}. To provide a comprehensive view of company emission objectives, we propose an approach  to source these metrics from company public disclosures. Without automation, curating these metrics manually is a labor-intensive process that requires combing through lengthy corporate sustainability disclosures that often do not follow a standard format. Furthermore, the resulting dataset needs to be validated thoroughly by Subject Matter Experts (SMEs), further lengthening the time-to-market.
  We introduce the Climate Artificial Intelligence for Corporate Decarbonization Metrics Extraction (CAI) model and pipeline, a novel approach utilizing Large Language Models (LLMs) to extract and validate linked metrics from corporate disclosures. We demonstrate that the process improves data collection efficiency and accuracy by automating data curation, validation, and metric scoring from public corporate disclosures. We further show that our results are agnostic to the choice of LLMs. This framework can be applied broadly to information extraction from textual data.

\end{abstract}

\ccsdesc[500]{Applied computing~Document searching}
\ccsdesc[300]{Mathematics of computing~Nonparametric statistics}
\ccsdesc[300]{Applied computing~Mathematics and statistics}


\keywords{Large Language Models, Corporate Transition Risk, Sustainability Analytics, Decarbonization, Temperature Alignment, Natural Language Processing, Finance}

\maketitle

\section{Introduction}

\subsection{Background}
To gain a comprehensive view of corporate emission targets, which are crucial for assessing a corporation's trajectory towards decarbonization, researchers typically turn to third-party aggregators, including the Science Based Targets initiative (SBTi) \cite{SBTiwebsite} and the Carbon Disclosure Project (CDP) \cite{CDPwebsite}. However, from our analysis, sizable percentage of companies (more than 30\% of the top 1,000 firms ranked by market capitalization across all marketable securities from developed, emerging, and frontier markets) do not disclose commitments data in a structured format through these third-party organizations. To help bridge this gap, we propose an information extraction framework to collect these metrics from company disclosures.

Information extraction from unstructured data is a well-known challenge in Natural Language Processing (NLP) \cite{infoextractionsurvey,wu2024learningextractstructuredentities,10032132}. Several areas of research have attempted to address this problem with respect to key-value metrics extraction highlighting limitations in the scalability and accuracy of the extraction models. \cite{seitl2024assessingqualityinformationextraction,limitationsunstructuredinformationextraction,liu2023lostmiddlelanguagemodels}. Information extraction in the context of corporate disclosure of climate risk metrics and objectives is challenging in particular, given a fast-evolving domain and heterogenous language. Several previous studies have explored this and related topics. In ChatClimate \cite{chatclimate} and ClimateQ\&A \cite{delacalzada2024climateqabridginggapclimate}, chatbots trained on IPCC (Intergovernmental Panel on Climate Change) documentation, utilized a RAG (Retrieval Augmented Generation) framework to identify relevant documents for a specific query.  ChatClimate's developers also created ClimateBERT, a climate-focused classifier derived from a distilled RoBERTa model, trained for climate language detection, sentiment analysis, and fact checking \cite{webersinke2022climatebertpretrainedlanguagemodel}. Additionally, MSCI/GARI \cite{mscisustainabilityinstitute} and Bank for International Settlements \cite{bisinnovationhub} studies employed LLMs to extract climate and sustainability metrics from texts using a RAG framework for precise text retrieval. While relevant, these studies are proof-of-concepts and do not produce production-ready results with high useable accuracies and stringent validation.

We contribute to the information extraction literature by demonstrating the effectiveness of a combination of techniques, including a fine-tuned RoBERTa model for enhanced text classification accuracy (instead of RAG), dynamic prompting, and domain-specific metric validation. We show that these techniques together achieve production-quality information extraction performance for the task of retrieving corporate emission targets from corporate disclosures.

In this paper, we also assess more traditional extraction methods like regular expression (RegEx) and BERT-based Question Answering, noting their cost-effectiveness but limited scalability. For example, Regex cannot comprehensively model and extract metrics of interest, and BERT struggles with multi-span extraction, hindering the capture of complete metric entities.

\subsection{Model Overview}


  \begin{figure*}

    \centering
    \includegraphics[width=\linewidth]{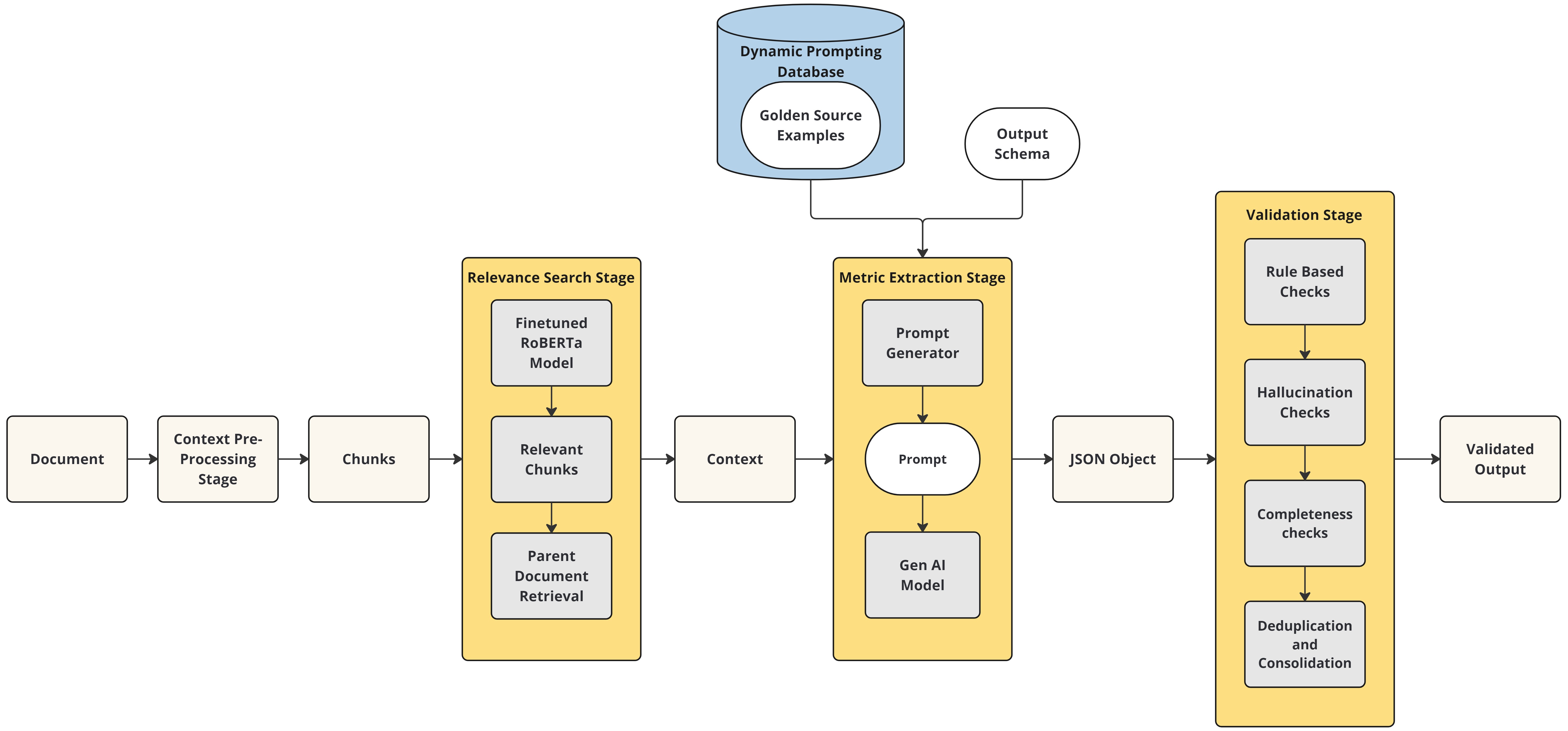}
    \caption{CAI model and pipeline end to end workflow.}
    \label{fig:workflow}
  \end{figure*}

In order to collect corporate carbon reduction commitments accurately, the CAI model utilizes 4 stages: context processing and chunking, relevant text search, metric extraction, and validation. All stages use NLP, and with the metric extraction step leveraging Large Language Models (LLM) and Generative AI (Gen AI). Stage 1 converts documents to text and breaks this text into smaller, consumable chunks for the pipeline. In the second phase of the process, a more advanced variant of the Bidirectional Encoder Representation Transformer (BERT), known as RoBERTa (Robustly Optimized BERT-Pretraining Approach), is employed. RoBERTa is built on top of BERT and has enhanced pretraining, such as larger training batches and dynamic word masking. RoBERTa, an encoder transformer-based language model, is tasked with the classification of text segments \cite{liu2019robertarobustlyoptimizedbert}. Specifically, it discerns whether these segments are relevant or not to the subject of carbon reduction commitments \cite{schimanski2023climatebertnetzerodetectingassessingnet}. This model enables a more nuanced understanding and categorization of the textual data, thereby enhancing the overall process by creating a search mechanism that improves context recall and the number of contexts to pass into the LLM in the later stage. The model casts a wide net and searches for all references of carbon reduction commitments, either corporate-wide or non-corporate wide, such as at the subsidiary level or country level. This was intentionally done to ensure that relevant texts were captured. Stage 3 is the metrics extraction layer which leverages an LLM, which gets fed in relevant blocks of text from stage 2 and extracts structured metrics from the input given a bespoke prompt. In the final stage, this output is then post-processed by transforming, validating and deduplicating data points by company. The end-to-end workflow of the model is depicted in Figure \ref{fig:workflow}.

\subsection{Data}

Public corporate sustainability and annual reports are used as inputs to the process. Corporate sustainability reports encapsulate initiatives that a firm is planning or has accomplished concerning sustainability and climate change. These documents typically encompass corporate commitments. However, a significant challenge arises as many firms do not publish sustainability reports regularly or at all. In such instances, annual reports are utilized for extraction. These documents usually contain a section on sustainability and corporate targets, while the remaining information is disregarded for the purpose of this model.
Each report is stored and tagged with metadata, including the corporate company identifier, name, report type (either annual or sustainability), and report publication year. Each input is converted from PDF to text and segmented into chunks for ingestion into the model.

\section{Methodology}

\subsection{Problem statement}

The objective of this model is to enhance and expedite the process of curating and confidence scoring corporate carbon reduction commitments. Before the advent of LLMs and Generative AI, this task was laborious, requiring significant manual effort for data collection and validation by SMEs \cite{ciravegna2001challenges}. With the progress of Gen AI, we are introducing a novel approach to factual data extraction, coupled with domain-specific validation checks, enabling us to curate data from unstructured text at scale.

To extract data from text, earlier approaches involved using RegEx to identify patterns in text. However, this is not a scalable approach, as most company disclosures lack a consistent language format or template for pattern matching. Regex provided an effective approach when employed to extract commitments from formulaic language, for example in SBTi disclosures which typically follow the format: "We plan to reduce [target type] emissions for [scope] by [target percent] by [target year] from base year [base year]." This method fell short when faced with more complex language structures.

An alternative approach involved using BERT Question Answering, a method that extracts the segment of text from an input that answers a specific question. While this approach could extract metrics from the input text, it was unable to extract values in a structured format like JSON. Moreover, individual queries had to be executed to extract each metric, such as base year or target year. Although functional, this process is challenging to scale as a unique question for each metric needs to be added to the training data, and linking disjoint metrics can become problematic. For instance, in the text "\ldots we plan to reduce Scope 1\&2 emissions by 50\% by 2030 and Scope 3 emission by 30\% by 2035 \ldots", running a query for the target year would yield the metrics 2030 and 2035 for each query, but associating the metrics with either scope 1 and 2 or 3 can become convoluted. Additionally, these encoder question answering models lack the ability to pay attention to multiple spans, which restricts them to extracting only a single answer from a block of text. Furthermore, this approach extracts only verbatim text, and the base model is required to be trained on a large number of human labelled data points. For example, if a sample target wording is described as " \ldots reduce absolute scope 1 and 2 emissions by 30\% and scope 3 emissions by 20\% by 2030 from 2015 \ldots", two commitments should be extracted as separate entities: both the scope 1+2 and the scope 3. An encoder model will extract the text as it is, leaving the question of how to extract the KPIs as distinct entities unanswered. 

In the case of CAI, a commitment entry needs to be collected that contains a minimum of five values in a structured format: target year, base year, target reduction percent, target type, and scope. In order to address this problem, a transformer architecture model is used. These models leverage convolutional neural networks which employ an encoder-decoder pattern \cite{vaswani2023attentionneed}. This architecture is considered the state-of-the-art in NLP. An autoencoder model \cite{investopedia}, or simply encoder model, would not be capable of extracting all these values, due to its non-generative nature and its core ability to simply learn an embedding representation of the input text. Generative AI models provide the capacity to perform multi-span attention and generate new texts by computing the likelihood of the next token given the previously seen tokens. This process is performed until a stop signal is predicted. Multi-span attention allows the LLM to focus on different relationships within an input context in parallel. These models are capable of few-shot learning, or the ability to use a few labeled examples to help generate the final result, as opposed to re-training the entire model \cite{dang2022promptopportunitieschallengeszero}. Moreover, generative AI models can be adapted to specific use-cases by prompting, allowing for the provision of bespoke or domain-specific information to the LLM at runtime without the need for fine-tuning. From the example above, Gen AI models would be able to curate each commitment in a structured format, such as in JSON, as the following:

\colorlet{punct}{red!50!black}
\definecolor{background}{HTML}{EEEEEE}
\definecolor{delim}{RGB}{20,105,176}
\colorlet{numb}{magenta!50!black}

\lstdefinelanguage{json}{
    basicstyle=\normalfont\ttfamily,
    numbers=left,
    numberstyle=\scriptsize,
    stepnumber=1,
    numbersep=8pt,
    showstringspaces=false,
    breaklines=true,
    frame=lines,
    backgroundcolor=\color{background},
    literate=
     *{0}{{{\color{numb}0}}}{1}
      {1}{{{\color{numb}1}}}{1}
      {2}{{{\color{numb}2}}}{1}
      {3}{{{\color{numb}3}}}{1}
      {4}{{{\color{numb}4}}}{1}
      {5}{{{\color{numb}5}}}{1}
      {6}{{{\color{numb}6}}}{1}
      {7}{{{\color{numb}7}}}{1}
      {8}{{{\color{numb}8}}}{1}
      {9}{{{\color{numb}9}}}{1}
      {:}{{{\color{punct}{:}}}}{1}
      {,}{{{\color{punct}{,}}}}{1}
      {\{}{{{\color{delim}{\{}}}}{1}
      {\}}{{{\color{delim}{\}}}}}{1}
      {[}{{{\color{delim}{[}}}}{1}
      {]}{{{\color{delim}{]}}}}{1},
}

\begin{lstlisting}[language=json,firstnumber=1]
[
  {
    "target_year": "2030",
    "base_year": "2015",
    "target_percent": "30%",
    "target_type": "absolute",
    "scope": "12"
  },
  {
    "target_year": "2030",
    "base_year": "2015",
    "target_percent": "20%",
    "target_type": "absolute",
    "scope": "3"
  }
]
\end{lstlisting}

The CAI model is architected to function sequentially, utilizing the output from one stage as the input for the next.  In the following, we describe the four stages of the CAI model and pipeline: Preprocessing, Relevance Search, Metrics Extraction, and Validation and Post Processing.

\subsection{Preprocessing}
The model accepts inputs in the form of PDF or text documents, which must be transformed into a format that the program can process. This involves converting the documents into plain text, cleaning the text, and segmenting it into manageable blocks or contexts.

This preprocessing phase is crucial as the model's relevance search and extraction mechanisms require a specific input structure and context size. Post conversion of a document to text, the text undergoes a cleaning process and is then divided into smaller contexts. The cleaning process involves normalizing special characters and spaces and substituting unusual characters like Unicode quotations with their ASCII counterparts.
The contexts are chunked using a sliding window algorithm with an overlap window, a strategy employed to prevent information loss that could occur with arbitrary segmentation methods. The overlapping of chunks ensures that relevant portions of the text are present in multiple n-gram chunks, thereby enhancing the recall and precision of relevant contexts during the search phase. Each chunk is approximately 80 words in length, with an overlap of 20 words on both sides. The optimal context window was evaluated using multiple window and overlap sizes with the above combination yielding the highest recall (see \autoref{sec:chunk} for details). Additionally, it is beneficial to keep the context chunks smaller to avoid LLM input and output token size constraints. This specific window length is also chosen to prevent extraneous information in the context from skewing the relevance classification in the subsequent stage. This hyperparameter choice also allows potentially relevant text to be at the start or end of the context, due to LLM's performing poorly on extraction tasks when the answer in the middle of the context \cite{liu2023lostmiddlelanguagemodels}. Furthermore, all preprocessed documents are cached for utilization in the following stage.

\subsection{Relevance Search}

In many Gen AI applications, a pure RAG based approach is applied for searching relevant text where the top $K$ most relevant context chunks are selected from a vector database based on similarity. In our initial testing, this approach did not yield a high recall and produced results with many false positives. This can be due to the nuanced and heterogenous nature of commitments language within corporate disclosures. Additionally, the pure RAG approach did not allow ease of finetuning the base vector embedding function. With these considerations, we opted for finetuning a RoBERTa based model which allowed for more control on the relevance search stage of the CAI model.

We finetuned the bias and classification layers of the base RoBERTa model to classify a context as either relevant or irrelevant by assessing whether a context includes information about corporate carbon reduction commitments. The model is a binary classifier, where only relevant contexts are fed into the subsequent stage. The RoBERTa model has a maximum text window size of 512 tokens, where a token is 1-1.5 words. We limit the context window to 80 words to enhance the inbuilt attention mechanism of the model to produce a more reliable classification. Since BERT-based encoder models requires that all inputs in a batch are of the same length, each input context is padded with empty strings to the maximum length of the contexts in the input set. The base model was trained on a dataset specifically curated with language from corporate carbon reduction commitments. This dataset comprised of over 1,000 data points, with a distribution of 60\% classified as relevant and the remaining 40\% classified as not relevant and a train-test split of 80/20 was used for finetuning. The computational training process was executed on a singular NVIDIA T4 GPU on GCP.

The RoBERTa model, in its original state, is pre-trained, signifying that it has undergone training via next word prediction and masked language model techniques \cite{liu2019robertarobustlyoptimizedbert}. It is crucial that this foundational functionality is preserved post fine-tuning. This is to circumvent a phenomenon known as catastrophic forgetting \cite{luo2024empiricalstudycatastrophicforgetting}, where the weights of the fine-tuned model are entirely superseded by the fine-tuning process, resulting in a complete loss of the initial base model state.
To ensure the preservation of the base model's knowledge during fine-tuning, a two-stage approach was employed. Initially, all layers, excluding the classification layer, were frozen, allowing only the bias and classifier weights layers to be retrained. This stage facilitated the model's adaptation to the new training data and domain. Subsequently, all layers were unfrozen, and the entire network was trained using a reduced learning rate. This second stage, by employing a lower learning rate, enhances the likelihood of reaching a global minimum in the gradient descent minimization function. This two-stage process ensures a balance between new knowledge acquisition and preservation of the base model's pre-existing knowledge.
Throughout the fine-tuning process, both the training and test losses were monitored to ensure a monotonically decreasing trend. This consistent decrease in loss verifies that the model continues to improve after each evaluation on the test hold-out-set. Following the fine-tuning process, the model achieved an f1-score, precision, and recall of 99\%.

The relevant contexts from the relevance search engine are then padded with the previous and the subsequent context from the document cache to build a larger context window. The previous and subsequent context do not need to be classified as relevant. This process is referred to as Parent Document Retrieval \cite{pdr}. The objective of this technique is to augment the relevant context with additional information that can be beneficial for extraction.

\subsection{Metrics Extraction}
The third stage of the pipeline is the metrics extraction engine. For each enriched relevant context, a bespoke prompt needs to be created to generate the most accurate extraction. This prompt employs an in-context-learning and dynamic prompting approach to both reduce hallucination and improve model extraction \cite{dang2022promptopportunitieschallengeszero}. Dynamic prompting \cite{su2022selectiveannotationmakeslanguage} is explained in detail in \autoref{sec:prompt}. After the prompt is formed by the dynamic prompting method, it is passed through the LLM, and a structured JSON response is returned.
To ensure the collection of a commitment for a given company and verify that the commitment is at the corporate level, two additional prompts are employed. These prompts classify whether the entity name matches the company name and determine if the target wording indicates a corporate-wide level commitment (as opposed to country-level, subsidiary-level, etc.). These basic prompts leverage in-context learning and few-shot learning to classify inputs, thereby reducing the potential risk of incorrectly mapping commitments \cite{dang2022promptopportunitieschallengeszero}.

\subsection{Validation and Post Processing}

Given the possibility of hallucination from LLMs, it is imperative to conduct domain-specific validations \cite{xu2024hallucinationinevitableinnatelimitation}. The validation stage is divided into four phases: transformation, scoring, deduplication, and enrichment. The transformation phase involves converting the JSON output for each context into a format suitable for later stage consumption. Next, the validation phase conducts rule-based data checks, completeness checks, and hallucination checks to generate a final confidence score for the record. The confidence score is a metric that helps us assess the confidence of a commitment record. The records are then arranged in descending order based on the confidence score computed in the previous step.

In some corporate disclosures,  the same emission targets are disclosed in different sections of their disclosures. The deduplication process utilizes a custom similarity scoring logic detailed below to identify commitments from the same company that exhibit the highest similarity to each other, thereby consolidating or eliminating duplicate records. To deduplicate, embeddings are first generated for the target wording, sub-context, and entity name attributes extracted from the context. These embeddings are in turn used to assess semantic similarity between other commitment objects to perform deduplication. For each company commitment, a similarity score is computed by averaging the cosine similarities between the embeddings of the target wordings, sub-contexts, and entity names, and the exact match scores for the remaining attributes.
A commitment is deemed similar to another commitment if their match score exceeds 0.95. This threshold was selected after evaluating multiple thresholds, and it yielded the highest precision and recall compared to the other thresholds. Once a commitment matches another, the consolidation and deduplication process commence.

Initially, similar commitments are consolidated. For instance, if commitment $A$ lacks an emission scope but contains all the other attributes present in commitment $B$, the scope from commitment $A$ is populated with the scope from commitment $B$. To select the optimal commitment from the set of similar commitments, a majority vote logic is employed to choose the best attributes. The rule-based and completeness validation calculators are re-run to capture the changes that were made due to metric consolidation. Any data points that have the same metrics are grouped and the record with the highest confidence score is preserved with the duplicate records getting dropped. A debug file of all the records before consolidation and deduplication is stored.

The final process of the validation phase is error code assignment. An entity match flag and emission boundary (corporate-wide vs non-corporate-wide) categories are generated for each record and error codes are applied to records that have validation infractions.

The CAI model's output reasonableness is evaluated by ground truth comparisons and extraction of target wording and sub-context from the input.
Ground truth testing, such as SBTi benchmarking, is utilized to determine if the commitments generated by the CAI model align with or resemble those from a recognized source like SBTi.
The CAI process also extracts two additional values from the input context: the target wording and sub-context. These values offer insights into the commitment's background and the commitment itself. This information, extracted from the LLM via prompting, can be filtered and/or reviewed for commitment relevance. For example, a sample target wording may be "Net Zero emissions" and the sub-context "We aim to reach Net Zero emissions across all emissions by 2050." These target wording and sub-context are considered relevant. On the contrary, a target wording of "food waste" and sub-context "Zero food waste for manufactured product" would be deemed irrelevant The target wording is classified either by emissions or non-emissions, so irrelevant records can be filtered out by this flag.

\section{Base Model}

The Google Vertex AI PaLM (Pathways Language Model) 2 for Text model suite \cite{anil2023palm2technicalreport} is leveraged as a base model and is provided as a managed service from GCP (Google Cloud Platform).
Text-bison is a foundational model that is optimized for various natural language task such as entity and information extraction. The model takes a max input of 8,192 tokens and returns a result with a maximum of 1,024 tokens. The text-bison model is available via an API call on GCP, and model parameters can be tweaked as necessary for use-case optimization.

The CAI model and pipeline was also evaluated using Open AI's GPT-3.5 and 4 models that are hosted on Azure ML and comparable results were yielded through a series of performance evaluation tests. Further detail on these tests is described in \autoref{sec:performance}. Overall, the CAI model and pipeline are robust against the choice of LLM base model. All performance metrics and statistics within this paper are based on the Vertex AI modelling suite, particularly the text-bison model, unless explicitly specified.

\section{Prompt Engineering}
\label{sec:prompt}
\subsection{Prompt Instruction and Schema}
In the metrics extraction pipeline, prompt engineering is employed to construct custom prompts that are contingent on the input context. Grounded examples are utilized to regulate and limit the knowledge of the LLM. The LLM is explicitly instructed to extract information solely from the user input, disregarding all other knowledge, thereby grounding its extraction based solely on the inputs, not the model's "memory". This is crucial as the purpose of the CAI model is to produce factual extractions. The prompt is further enhanced with an output schema structure, and strict adherence to it is enforced. If the LLM is unable to extract a specific attribute from the schema from the input, such as the target base year, it is instructed to default returning "NO\_ANSWER" for that attribute. The CAI model uses a modified RAG approach \cite{gao2024retrievalaugmentedgenerationlargelanguage} which uses the input context as the input query and the top $K$ examples as the context augmentation.
\subsection{Dynamic Prompting}
The subsequent challenge lies in identifying the most suitable examples for the input context to enhance extraction quality. This can be accomplished by employing dynamic prompting, a technique that utilizes a similarity function to identify an example most similar to the input text. This is achieved by comparing the input text to all the examples in the golden source example database. LLM instruction, schema enforcement, and dynamic prompting are implemented to mitigate hallucination and enhance model extraction.
The CAI pipeline incorporates 71 ground truth examples that were manually gathered and vetted by a team of SMEs. At the time of prompt creation, the most similar examples are selected from the ground truth examples between the input context and the example sub-context. The sub-context is a segment of the full context that contains the commitment language, such as "reduce scope 1 and 2 emissions by 30\% by 2030 from FY20."
Through this process, the model is provided with examples that have the most similar language to the input context, such as context and sub-context both including terms like "Net Zero" or "absolute scope 1 and 2 emissions". These examples also prime the LLM to better locate the sub-context for the input context that can aid model attention and can later be validated. Given these similar examples, a prompt is generated using the Kor python library which formats the prompt with a schema and structure to improve prompt readability and interpretation by the LLM \cite{bronzini2024glittergoldderivingstructured,ni2023chatreportdemocratizingsustainabilitydisclosure}.

\section{Evaluation Metrics and Methodology}

\begin{table*}
  \caption{Model Evaluation Results Against Train and Evaluation Out-of-Sample SBTi Datasets}
  \label{tab:sbti_results}
  \begin{tabular}{c c c c c c}
    \toprule
    \textbf{Test Type} & \textbf{Num company Reports} & \textbf{Num Unique Commitments} & \textbf{Accuracy} & \textbf{Recall} & \textbf{Precision} \\ \midrule
    Train       & 35             & 102               & 95\%       & 96\%      & 90\%        \\ 
    Validation     & 11             & 42               & 100\%       & 100\%      & 100\%       \\ \bottomrule
  \end{tabular}
\end{table*}

\subsection{Overview of Testing Methods and Results}
The CAI testing process is comprised of several components that include confidence scoring, accuracy measurement against a ground truth golden dataset and manual review with assisted tooling to support the review process.
Confidence scores are calculated by averaging 3 separate metrics: a rule-based validation score, completeness score and hallucination score. Additional to this score, descriptive error codes are also generated depending on validation check breach to assist manual review, such as "invalid target year" if the target year is not greater than the base year for example.

\subsection{Model Accuracy Testing}

The validation of the CAI model's accuracy is conducted by aligning its outputs with a verified benchmark set, which is either assembled by SMEs or obtained from an external source such as SBTi. A company's commitment object is deemed accurate if it matches the benchmark object in all attributes, including target year, target percent, base year, target type, and scope. For instance, if a document contains five benchmark commitment objects \{\textbf{A}, \textbf{B}, \textbf{C}, \textbf{D}, \textbf{E}\} and the CAI process yields \{\textbf{A}, \textbf{B}, \textbf{C}, \textbf{E}, \textbf{F}\}, the accuracy, precision, and recall rate would be 80\%. However, if the commitment objects identified by CAI \{\textbf{A}, \textbf{B}, \textbf{C}\} are of high confidence (indicated by a 100\% confidence score), the recall and precision rates are considered to be 100\% but would be 60\% total recall and 100\% precision. These metrics are calculated at the document level, with the confidence score serving as the metric for the commitment itself. The primary metric of focus is total recall, aiming for the CAI model and pipeline to identify over 95\% of the commitments within a document. The threshold of 95\% was chosen as a baseline confidence level based on practicality, due to some metrics being mentioned in tables or images that may be difficult to extract out. The secondary metric, precision, is also crucial as it reduces the likelihood of extracting incorrect or irrelevant data.

\subsection{Sensitivity Analysis}

Sensitivity analysis was conducted for context chunking size, prompt, and hyperparameter tuning of the models. Each of these approaches can significantly impact the output result, necessitating rigorous testing and analysis.

\subsubsection{Context Chunking Size Tuning}
\label{sec:chunk}
\begin{figure}[h]
  \centering
  \includegraphics[width=\linewidth]{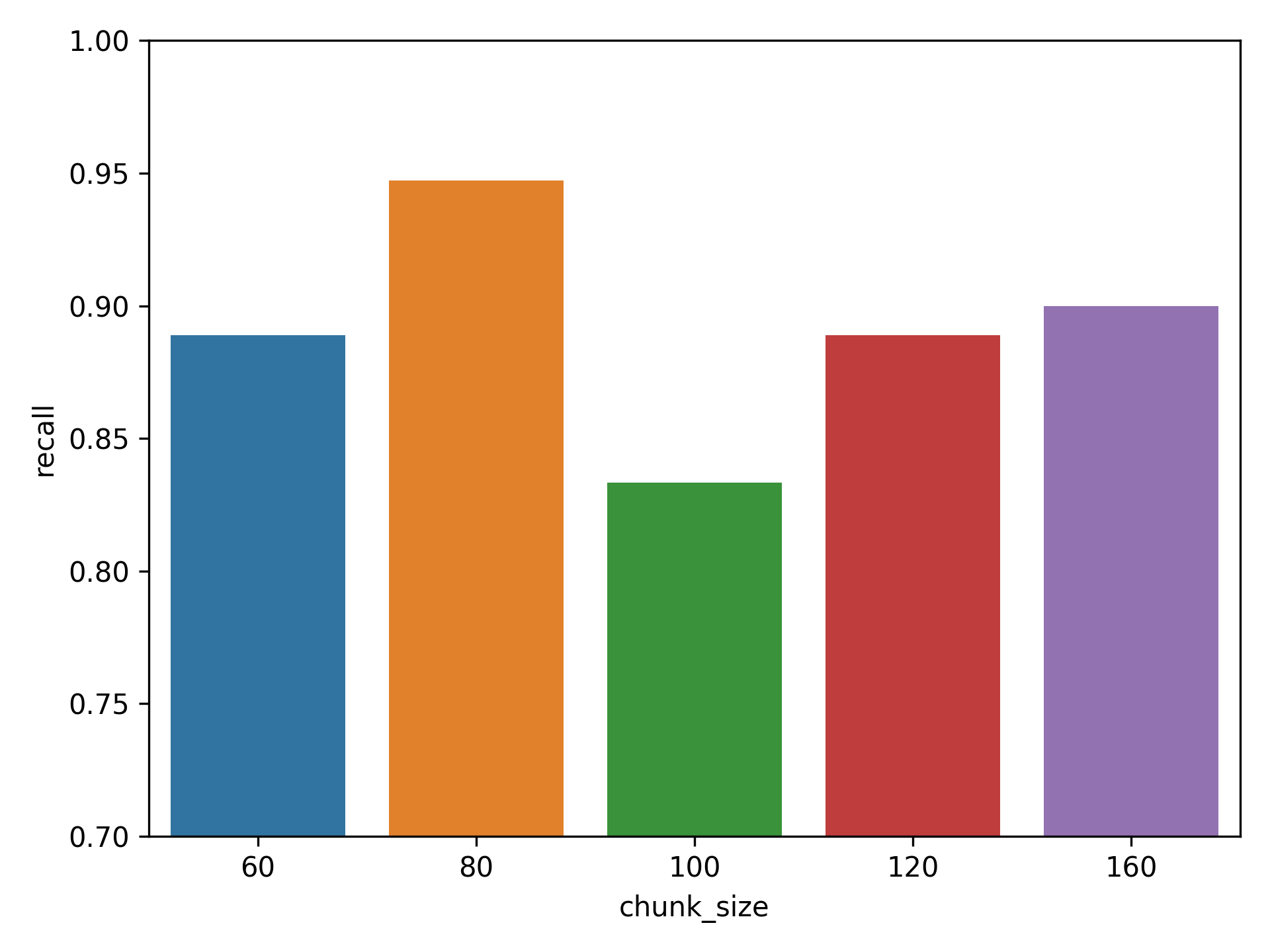}
  \caption{Chunk size experiment recalls.}
  \label{fig:chunk_size_experiment_recalls_}
\end{figure}

The tuning of context chunking size is particularly crucial for the RoBERTa model classification. The size of the context chunk can significantly influence the model's performance. If the chunk size is too small, there is a risk of losing sentence context. Conversely, if the chunk size is too large, the classification may be biased towards unnecessary tokens in the context. In this study, the input context chunk size was evaluated with sizes of 60, 80, 100, 120, and 160 words with an overlap window of 1/4 of the total word count. The $Y$ variable for this test was the recall from the relevant chunks that were yielded from the classification model given the input context chunk size. The recall could be calculated from the relevant chunks by checking how many of the chunks contained the commitment metrics from a curated golden source dataset. The optimal word count was found to be 80 with an overlap window of 20, which yielded the highest recall of 95\% against 43 examples. The breakdown of the context chunking test can be referred to in Figure \ref{fig:chunk_size_experiment_recalls_}.

\subsubsection{Prompt Tuning}

\begin{figure}[h]
  \centering
  \includegraphics[width=\linewidth]{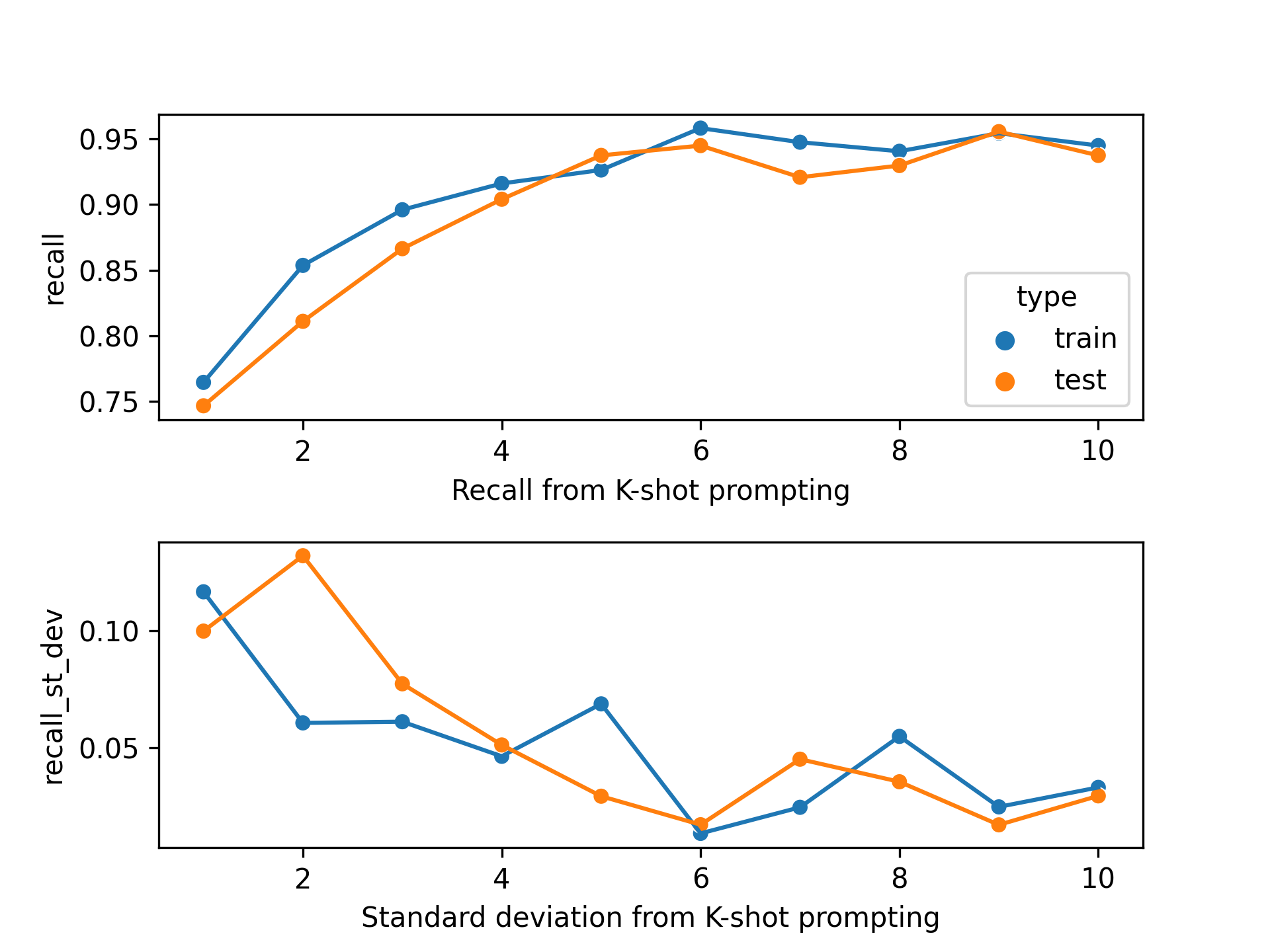}
  \caption{Prompt K-shot tuning.}
  \label{fig:k-shot_recall}
\end{figure}

The next phase of sensitivity testing focused on the prompt itself. The prompt can comprise of the following components: an instruction that explains the task and enforces bounds and rules, and examples, or "shots" that provide an output example result for the LLM to build its answer off. An optimal top-K number of examples was identified by running the test for k-shot values for 1 to 10 examples inclusive. There were 6 random cross validation samples run for each K-shot. This test was completed by breaking the ground truth example set into a train and test set, using a 70/30 split from 50 out-of-sample examples. The dataset was split by grouping on context, to ensure a specific context did not exist in both sets. The purpose of breaking the dataset into train and test split is to validate whether hyperparameter choices are consistent, even if the example contexts change slightly. The prompt with six examples yielded accuracy and recall of 97\% in the train set and 95\% for the test set, the highest amongst the tests with there being a diminishing marginal recall after 6 shots. The k-shot prompt tuning output can be viewed in Figure \ref{fig:k-shot_recall} for the Vertex AI text-bison model.

\subsubsection{LLM Hyperparameter Tuning}

The final approach involved modifying the input hyperparameters of the LLM itself. The primary parameters for tuning included temperature, top-K, and top-P. These parameters were all changed together due to their interrelated nature. Two tests were performed: one where the temperature, top-P, and top-K were all kept at their defaults (respectively 0.7, 0.95, 40), and another where the temperature was kept at 0, top-P kept at 0, and top-K kept as 1. The second test proved to produce more reliable, reproducible results that could be evaluated consistently with an accuracy and recall of 95\% for out of sample data points.

\begin{table*}
  \caption{Gen AI Model Performance Comparison}
  \label{tab:results}
  \begin{tabular}{c c c c}
    \toprule
    \textbf{Model}            & \textbf{Total Recall} & \textbf{High Scored Data Precision} & \textbf{High Scored Data Recall} \\ \midrule
    Vertex AI text-bison         & 92\%         & 100\%                & 45\%               \\ 
    Azure Open AI GPT-3.5 Turbo Instruct & 93\%         & 100\%                & 50\%               \\
    Azure Open AI GPT-4 (gpt-4-0613)   & 97\%         & 100\%                & 58\%               \\ \bottomrule
  \end{tabular}
\end{table*}

\section{Gen-AI Model Performance Comparison}
\label{sec:performance}

A comparative analysis was conducted between two Gen AI LLMs to evaluate their output recall, and precision. The models were assessed using an input set of 10 companies from the SBTi with 26 unique commitments.
The default LLM used for this comparison was Vertex AI’s text-bison model. As a point of comparison, the OpenAI Chat GPT-3.5 Turbo Instruct model was also utilized. For vector similarity search, the Open AI text-embedding-ada-002 was compared against the Vertex AI embedding function text-embedding-gecko. The Instruct model from the GPT suite was leveraged as it provided the best opportunity to run an apples-to-apples test against the extraction capabilities of the text-bison model since their training is similar. All prompt instruction and hyper-parameters for each model was kept the same. The Chat GPT-3.5 Turbo Instruct model demonstrated the ability to exceed the recall and precision of the text-bison model, despite requiring fewer in-context examples.
For the text-bison model, the generation of a prompt with 5 to 8 examples resulted in the highest recall and precision. Specifically, this model achieved a total recall of 92\%, recall and precision against records with no error codes of 45\% and 100\% respectively mentioned in Table \ref{tab:results}.
On the other hand, the Chat GPT-3.5 Turbo Instruct model demonstrated a remarkable efficiency. Depending on the input token size, it required only 1 to 5 examples to yield a total recall of 93\% and recall and precision against records with no error codes of 50\% and 100\% respectively, which suggests that the Chat GPT-3.5 Turbo Instruct model may offer superior performance with less data, highlighting its potential for efficient and accurate language generation tasks.
Another comparison between the text-bison and the GPT-3.5 Turbo Instruct model results from the CAI model and pipeline was passing in text with or without carriage character stripping. The text-bison model demonstrated a 70\% total recall with recall and precision against records with no error codes of 20\% and 100\% respectively when carriage characters were not stripped, as compared with the results above where they were stripped. The GPT-3.5 Turbo Instruct improved overall recall to 95\% and other metrics remained the same from the results above where they were stripped.

The evaluation of the test data against the GPT-4 model, which boasts a staggering one trillion parameters—a significant leap from the 175 billion parameters of its predecessor, GPT-3.5 Turbo Instruct—revealed notable results. GPT-4’s inclusion of a random seed parameter enhances reproducibility, a critical aspect in research. The model demonstrated exceptional performance, with a total recall rate greater than 95\%. Furthermore, when assessing records devoid of error codes, GPT-4 achieved a recall of 58\% and a precision rate of 100\%, indicating a high level of accuracy in its predictions.

\section{Results}

The output of the CAI model is benchmarked against commitment data from SBTi. Given that SBTi commitments are verified by a third party, it is generally regarded as the industry standard for target disclosures and therefore they possess a higher level of confidence on average compared to manually extracted commitments.
For benchmarking, SBTi commitments for 35 companies are compiled, along with the reports containing the targets reported by SBTi as shown in Table \ref{tab:sbti_results}. It is crucial that these reports include the targets published by SBTi to facilitate accurate extraction benchmarking.
In total, there were 102 targets from SBTi in the training set and 42 targets in the test set. These datasets were defined by randomly choosing companies with quantitative targets from the SBTi dataset and using a 70/30 train test split to build each dataset. It is important to note that SBTi companies were used for testing if their respective targets exist in either an annual report or sustainability report to effectively evaluate the CAI model performance. If a random company chosen from SBTi  does not meet the criteria, it is ignored in subsequent draws and a new company is sampled. The reconciliation of the training set yielded an accuracy of 95\%, a recall of 96\%, and a precision of 90\%. 

The test set reconciliation, on the other hand, achieved a perfect score with 100\% accuracy, recall, and precision. It is noteworthy that the CAI model and pipeline collects more targets than SBTi for enhanced coverage. One key point to address on discrepancies during reconciliation is that SBTi mandates a 24-month review for commitments \cite{SBTiwebsite,sbtifaqs}. During this period, firms' targets can be labeled as committed, but quantitative values may not be approved yet. The CAI model can expedite time to market by extracting data points during these approval processes. Additionally, we expect the model performance to increase if we curate the target set ourselves as opposed to using a third-party vendor due to more control in the extraction process, such as expanding further KPI curation and enforcing specific output validation standards. 

With respect to evaluating other Gen AI models through the CAI pipeline, all models described in Table \ref{tab:results} performed well at extracting numerical KPIs like target percent or target year, but Vertex AI struggled the most with extracting scope and target type, which can be more ambiguous metrics to capture. GPT 3.5 performed a bit better than Vertex AI, but also struggled on the same issues. GPT-4 was able to correctly extract target types such as "net zero" or "intensity" as well as accurately translate scope specific language such as equating "own operations" with "scope 1+2".

This study leverages the CAI model to enhance the coverage of the top 1,000 firms ranked by market capitalization across all marketable securities from developed, emerging, and frontier markets. Our findings indicate a significant improvement in coverage, increasing from approximately 68\% when utilizing solely SBTi and CDP data, to 78\% with the inclusion of 92 additional firms with public disclosures. Future efforts will focus on further bridging this gap by incorporating an increasing number of reports from additional firms into the model pipeline.

\section{Conclusions}

Leveraging the latest research in Gen AI and NLP, we built an automated data pipeline that is specifically tuned to extracting and validating corporate carbon reduction commitments from public corporate disclosures. We employed a scalable process that accurately curates climate-specific structured data from unstructured source using a combination of techniques spanning context chunking, relevance search, metric extraction, and metric validation. Using a combination of encoder-based search with various prompting strategies such as Parent Document Retrieval and Dynamic Prompting, we were able to efficiently curate data previously not available at scale.

There are still several areas that require further research, such as dynamic top-K shot example selection for prompting or LLM-based QA against the output yielded to improve the outputs of the model.

\section{Acknowledgments}

The views expressed here are those of the authors alone and not of BlackRock, Inc. 
We thank Dimitrios Vamvourellis for advice on LLM prompting strategies and evaluation metric creation.


\bibliographystyle{ACM-Reference-Format}

\bibliography{ACAI_research_paper}

\end{document}